\documentclass[twocolumn,showpacs,preprintnumbers,amsmath,amssymb,floatfix]{revtex4}
\usepackage[utf8]{inputenc}
\usepackage{amsmath}
\usepackage{subfigure}
\usepackage{array}
\usepackage{graphicx}
\usepackage{fancyheadings}
\usepackage{dcolumn}
\usepackage{bm}
\usepackage{graphicx,color}

\def\be{\begin{equation}}
\def\ee{\end{equation}}
\def\ba{\begin{eqnarray}}
\def\ea{\end{eqnarray}}

\begin{document}

\title{The renormalized Jellium model of colloidal suspensions with multivalent counterions}

\author{Thiago E. Colla}
\email{colla@if.ufrgs.br}
\author{Yan Levin}
\email{levin@if.ufrgs.br}

\affiliation{
Instituto de F\'isica, Universidade Fedaral do Rio Grande do Sul, CP 15051, 91501-970 Porto Alegre, RS, Brazil.
}

\begin{abstract}

An extension of the renormalized Jellium model which allows to study colloidal suspensions containing trivalent counterions is proposed. 
The theory is based on a modified Poisson-Boltzmann equation which incorporates the effects of counterion correlations 
near the colloidal surfaces using a new boundary condition. 
The renormalized charges, the counterion density profiles, and osmotic pressures can be easily calculated using the modified 
renormalized Jellium model.
The results are compared with the ones obtained using the traditional Wigner-Seitz (WS) cell approximation also with a new boundary condition.  
We find that while
the thermodynamic functions obtained within the renormalized Jellium model are in a good agreement with their WS counterpart,
the effective charges predicted by the two theories can be significantly different.
\end{abstract}

\maketitle
\section{Introduction}
Strongly acidic or basic groups are often used to stabilize colloidal suspensions against flocculation and precipitation. The resulting electrostatic 
repulsion prevents the like-charged particles from coming sufficiently near one another where a short-range van der Waals attraction
can result in an irreversible association.
It is well known that the Poisson-Boltzmann (PB) theory describes accurately the properties  of such colloidal suspensions 
containing a symmetric 1:1 electrolyte \cite{1,2}. For these systems the correlations between small, strongly hydrated ions are weak, 
making the mean-field PB equation quasi-exact. This is no longer 
true when 
the counterions are  multivalent. In this case, electrostatic correlations between the ions can no longer be neglected 
and the mean-field PB equation looses its validity. 
Although the PB equation predicts very similar behaviors for suspensions containing either monovalent or multivalent counterions \cite{3}, both experiments and simulations show that this is not correct. Many interesting phenomena inherent to colloidal suspensions with multivalent counterions, such as colloidal charge reversal \cite{4,5,6,7}, like-charged colloidal attraction \cite{8}, or the reversal of the electrophoretic mobility \cite{9} can not be captured by the simple mean field approach. The description of these phenomena requires the use of more sophisticated and numerically more demanding approaches, such as the Monte Carlo simulations \cite{10}, the integral equations \cite{11}, or the density functional theories \cite{12}. 

To account for the electrostatic correlations in suspensions containing multivalent counterions, while preserving the simplicity of the PB theory, dos Santos \textit{et al.} proposed a Wigner-Seitz (WS) model with a modified boundary condition at the colloidal surface \cite{13,14}. The model is based on the Shklovskii's idea \cite{15} that the validity of the mean-field theory can be extended if the correlations between the condensed counterions are properly taken into account. 
In the context of the PB equation, this can be done using a modified boundary condition which must be satisfied by the mean-field potential at the colloidal surface. Using this new boundary condition, the PB equation can be solved inside the Wigner-Seitz cell to yield the ionic density profiles from which
the effective colloidal charge can be calculated using the Alexander prescription \cite{13,16}. 

The WS cell model is based on the
assumption that the liquid state structure of a colloidal suspension is not very different from a crystal one.  This allows one to avoid the complicated many-body problem by considering only the electrostatic interaction between one colloidal particle and its counterions and coions. As a further approximation, the geometry of the WS cell is taken to match the colloidal one (e. g. spherical). The electro-neutrality condition is imposed by requiring that the electric
field vanishes at the cell boundary.  Thus, within the cell model there no direct electrostatic interactions between the different colloidal particles.  Nevertheless, the effective  colloidal charge obtained by matching the linear and nonlinear solutions of the PB equation at the cell boundary is often
used within the  Derjaguin-Landau-Verwey-Overbeek (DLVO)  potential \cite{17} to account for particle-particle interaction.

The cell model should be
particularly appropriate for suspensions with large volume fractions of colloidal particles, while its validity for dilute systems is questionable. 
In this paper  we propose a more liquid-state-like point of view, which is particularly appropriate
for suspensions with low volume fraction of colloidal particles.  

Inside a suspension each colloidal particle feels the field produced
by other particles and their condensed counterions.   If we neglect the correlations between the colloidal particles
they, together with their condensed counterions, can be thought to provide a uniform background in which free microions move. 
The effective colloidal charge and the background charge must be determined self-consistently, resulting in a  Renormalized Jellium Model (RJM) \cite{18}. 
This theory is particularly appropriate for calculating the effective colloidal charges because  within the RJM, unlike in the
cell model,
the interaction potential between the colloidal particles has precisely the DLVO form \cite{18}.  
The RJM has been successfully used to calculate the effective charges and the 
structural and thermodynamic properties of colloidal suspensions containing 1:1 electrolyte \cite{17,19,20,21}.  
In this paper the RJM will be extended to account for the electrostatic correlations in 
suspensions with trivalent counterions.

The paper is organized as follows. In section II, the general aspects of the model will be outlined. In section III we will describe the RJM with a modified boundary condition.   In section IV the results of the theory will be presented. Finally, the conclusions and discussion will be given in section V.

\section{The Model}

We consider a system of colloidal particles of radius $a$ and (negative) charge $- Z_{bare} q$ (uniformly distributed on the surfaces), and counterions of radius $r_{c}$ and charge $\alpha q$, ($q$ is the charge of proton) inside an aqueous solution of volume $V$. We adopt a primitive model (PM) description in which the solvent is treated as a uniform continuum of dielectric constant $\epsilon$. The overall charge neutrality requires $N_{c} \alpha - N Z_{bare}=0$, where $N_{c}$ and $N$ are the particle numbers of counterions and colloids, respectively. The typical length scale that characterizes the system is the Bjerrum length, defined as  $\lambda_{b} \equiv \frac{\beta q^{2}}{\epsilon}$,  which is  $7.2$ \AA,  in water at room temperature. In order to maintain the simplicity of the model, image charge effects which can become non-trivial for multivalent counterions \cite{22}, are neglected within the primitive model approach adopted here --- colloidal particle has the same dielectric constant as the solvent.

Because of strong electrostatic interaction between the counterions and the colloidal particles, many of the counterions become condensed onto colloidal surface \cite{1,2}. A 'complex' composed of one colloidal particles with a layer of its condensed counterions can then be regarded as a single entity carrying an effective charge $Z_{eff}\ll Z_{bare}$ \cite{1,2}. The charge neutrality condition then becomes $\rho_{f}\alpha-Z_{eff}\rho=0$, where $\rho=N/V$, and $\rho_{f}$ is the number density of \textit{free} unassociated counterions. 

Traditionally colloidal suspensions have been modeled using a crystal-like approximation of a single macroion with its counterions inside a Wigner-Seitz cell, the
radius of which is determined by the volume fraction of colloidal particles.  Although this picture is appropriate at large concentrations  ---
when strong correlations between the charged particles lead to a crystal-like ordering --- it might not be appropriate for dilute suspension. 
To avoid the WS cell approximation we will instead use a liquid-state RJM. 

Suppose we fix one colloidal particle at the origin.   Far from this particle the counterion density profile will have a Boltzmann-like form, $\rho_{f}(r)=\rho_{f}e^{-\beta q \alpha \phi(r)}$, where $ \phi(r)$ is the mean electrostatic potential.  
Note that $\rho_f$ refers only to free, uncondensed, counterions.  The density of 
other colloidal particles and of their condensed counterions provides a uniform neutralizing background $\rho_{back}=\rho Z_{back}$.  
The mean electrostatic potential satisfies the Jellium-Poisson-Boltzmann (JPB) equation:
\begin{equation}
 \nabla^{2}\psi(r)=\dfrac{\kappa^{2}}{\alpha}\left(1-e^{-\alpha \psi(r)}\right)+\dfrac{Z_{bare}\lambda_{b}}{a^{2}}\delta(r-a),
\end{equation}
where $\psi(r)\equiv\beta q \phi(r)$ is the reduced potential, and $\kappa^{2}=4\pi\lambda_{b}\rho_{f}\alpha^{2}=4\pi\lambda_{b}\rho Z_{back} \alpha$ defines the inverse effective Debye screening length.  
The self-consistency condition requires that the effective charge, calculated from the far field solution of this equation, and
the background charge must have the same value, $Z_{back}=Z_{eff}$.  We should note that the  screening of the electrostatic potential in the far field 
is produced only by the free (uncondensed) ions. A similar behavior is implicit within in the cell model if one tries to define the effective
colloidal charge, as is done within the Alexander prescription \cite{16}.  However, this renormalization is less transparent within the cell model than 
within the Jellium formalism. 

Although the above mean-field equation works very well for monovalent ions $\alpha=1$, it becomes a rather poor approximation when suspension contains  multivalent counterions ($\alpha>1$). In such cases, the strong electrostatic correlations between the condensed counterions lead to significant
deviations from the PB theory.  The deviations can be so strong that they qualitative modify the behavior of suspensions containing multivalent counterions.  
In such suspensions,  one finds that the counterion condensation can become so strong as to reverse the sign of the effective colloidal charge.
Furthermore, addition of a multivalent electrolyte can result in attraction between like-charged
colloidal particles, thus destabilizing suspension against flocculation.  
 
To include the effects of the counterion correlations one can proceed in a number of different ways. One approach is to use a weighted-density functional theory \cite{23}
to account for the corrections to the mean-field electrostatic potential.  Another approach is to  
use the integral equations theory \cite{24}.  All of these methods, however, have their own drawbacks and are
significantly more computationally demanding than the simple WS cell PB theory or the RJM.

We note that the counterion
correlations are the strongest among the condensed counterions, since these ions are in the closest proximity of each other.  
Following Shklovskii we will, therefore, attempt to include the counterion correlations within the RJM using a
modified boundary condition at the colloidal surface.
The condensed counterions will be treated as a strongly correlated fluid --- a concentrated quasi-two dimensional plasma. On the other hand, in the bulk
the concentration of counterions is quite small, so that the correlations can be neglected and the mean field approach is still sufficient. 
By matching the two regimes, the JPB equation is recovered, but with a new boundary condition at the colloidal surface.
From now on, we will restrict our attention to trivalent counterions, $\alpha=3$.

\section{The Theory}

Strong electrostatic interactions between the colloidal particles and their counterions lead to counterion condensation.  
The condensed ions are in thermodynamic equilibrium with the free ions of suspension.
Close to the colloidal surface the counterion chemical potential is,
\begin{equation}
 \beta \mu_{sc}=\ln(\Lambda^{3}\rho_{sc})+\beta \mu_{c}+\beta q\alpha \phi(a+r_{c}),
\end{equation}
where $\Lambda$ is the de Broglie thermal wavelength and $\rho_{sc}$ is the course-grained density of condensed counterions. 
The correlational chemical potential $\mu_{c}$ is given by that of a two-dimensional one component plasma \cite{13,25}:
\begin{equation}
 \beta \mu_{c}=-1.65\Gamma+2.61\Gamma^{1/4}-0.26\log(\Gamma)-1.95,
\end{equation}
were $\Gamma\equiv\frac{\alpha^{3/2}\lambda_{b}\sqrt{Z_{bare}}}{2(a+r_{c})}$ is the plasma parameter. 
Far away, in the bulk solution, the chemical potential is well approximated using the mean field electrostatic potential,
\begin{equation}
 \beta \mu_{b}(r)= \ln[\Lambda^{3}\rho_{f}(r)]+\beta q \alpha \phi(r).
\end{equation}
The thermodynamic equilibrium between the condensed counterions and the free ions requires equality of Eqs. (2) and (4), from which follows
\begin{equation}
 \rho_{f}(r)=\rho_{sc} e^{\beta \mu_{c}}e^{-\beta q\alpha(\phi(r)-\phi(a+r_c))}.
\end{equation}
This equation correctly describes the density profile after a short distance $\delta$ from the colloidal surface. This cut-off distance delimits the region where the microion correlations are important and the mean field approximation breaks down. However, since the range of counterion correlations is quite small and the PB density profile vary smoothly, Eq. (5) can be extrapolated all the way to the colloidal surface. This simplification then results in a new boundary condition for the JPB equation at the colloidal surface,
\begin{equation}
 \rho_{f}(a+r_{c})=\rho_{sc} e^{\beta \mu_{c}}
\end{equation}
The value of $\rho_{sc}$ can be obtained using the strong coupling theory \cite{26} and coarse 
graining procedure \cite{13}.  We find
\begin{equation}
 \rho_{sc}=\dfrac{Z_{bare}^{2}\lambda_{b}}{8\pi(3.701)(a+r_{c})^{4}}.
\end{equation}
The strong dependence of the coarse grained density on the colloidal radius is a direct consequence of the contact theorem which states that the difference between the contact and the bulk density of counterions is proportional to the square of the electric field \cite{27}. 
Since the bulk counterion density is much lower than the counterion concentration at the colloidal surface, the theorem requires that the density near the
colloidal surface scale as $1/(a+r_{c})^{4}$ \cite{13} .
Together, Eqs. (3), (6) and (7) provide a new relation between the density profile at the colloidal surface and the bare colloidal charge $Z_{bare}$. This should be contrasted with the usual boundary condition for the PB equation, $\frac{d\phi(r)}{dr}\arrowvert_{r=a}=\frac{Z_{bare}\lambda_{b}}{a^{2}}$ which, however,
does not capture the strong counterion condensation resulting from electrostatic correlations at short distance from the colloidal surface.   

The calculation of the electrostatic potential from Eq. (1) is now quite straight forward. Far from the colloidal particle the electrostatic potential
has {\it exactly} the DLVO form,
\begin{equation}
\psi(r)=\frac{Z_{eff}\lambda_{b}e^{\kappa a} }{(1+\kappa a)}\frac{e^{-\kappa r}}{r}.
\end{equation}
Now suppose that we know $Z_{eff}=Z_{back}$, $r_{c}$, $a$, and the volume fraction $\eta=4\pi a^{3} \rho/3$.  Eq. (8) then provides the
electrostatic potential and the electric field far from the colloidal surface.  
Using these as the initial conditions, we can numerically integrate the JPB equation (1) to obtain the density $\rho_f(a+r_c)$.  Numerical integration  is
performed using the usual Runge-Kutta algorithm.  Eqs. (6) and
(7) can then be used to calculate the bare colloidal charge.  In practice, of course, we know the bare charge and would like to calculate $Z_{eff}$.  This can be
easily done by coupling the JPB solver with a root finding subroutine, such as the Newton–Raphson method.  For each $Z_{eff}$ there is a corresponding 
$Z_{bare}$.  The root finding subroutine allows us to efficiently search the values of  $Z_{eff}$ to find the one that corresponds to the given $Z_{bare}$.  

\section{Results}

\begin{figure}
\centering
\subfigure
{\includegraphics[width=4.5cm,height=4.0cm]{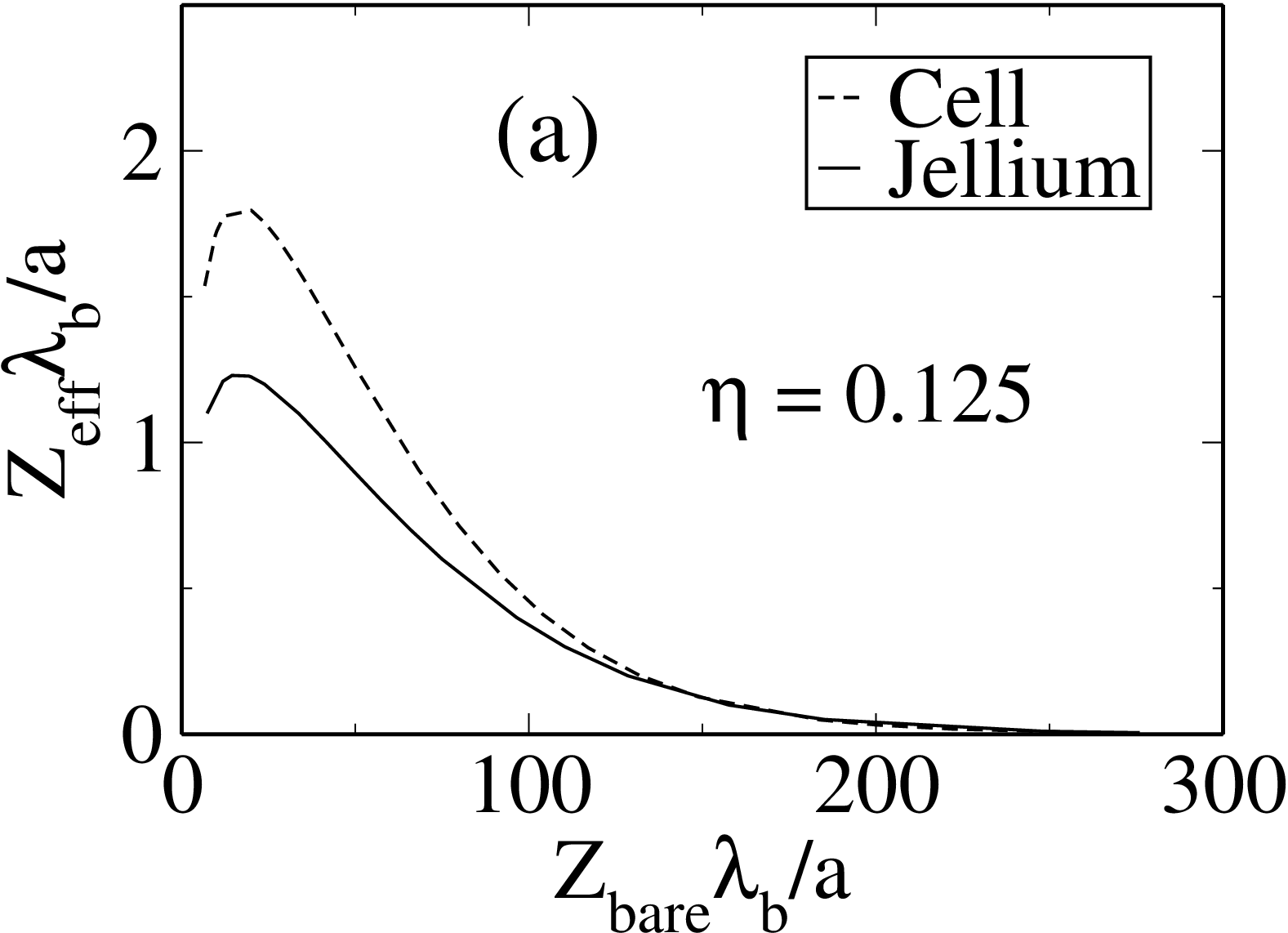}
\includegraphics[width=4.5cm,height=4.0cm]{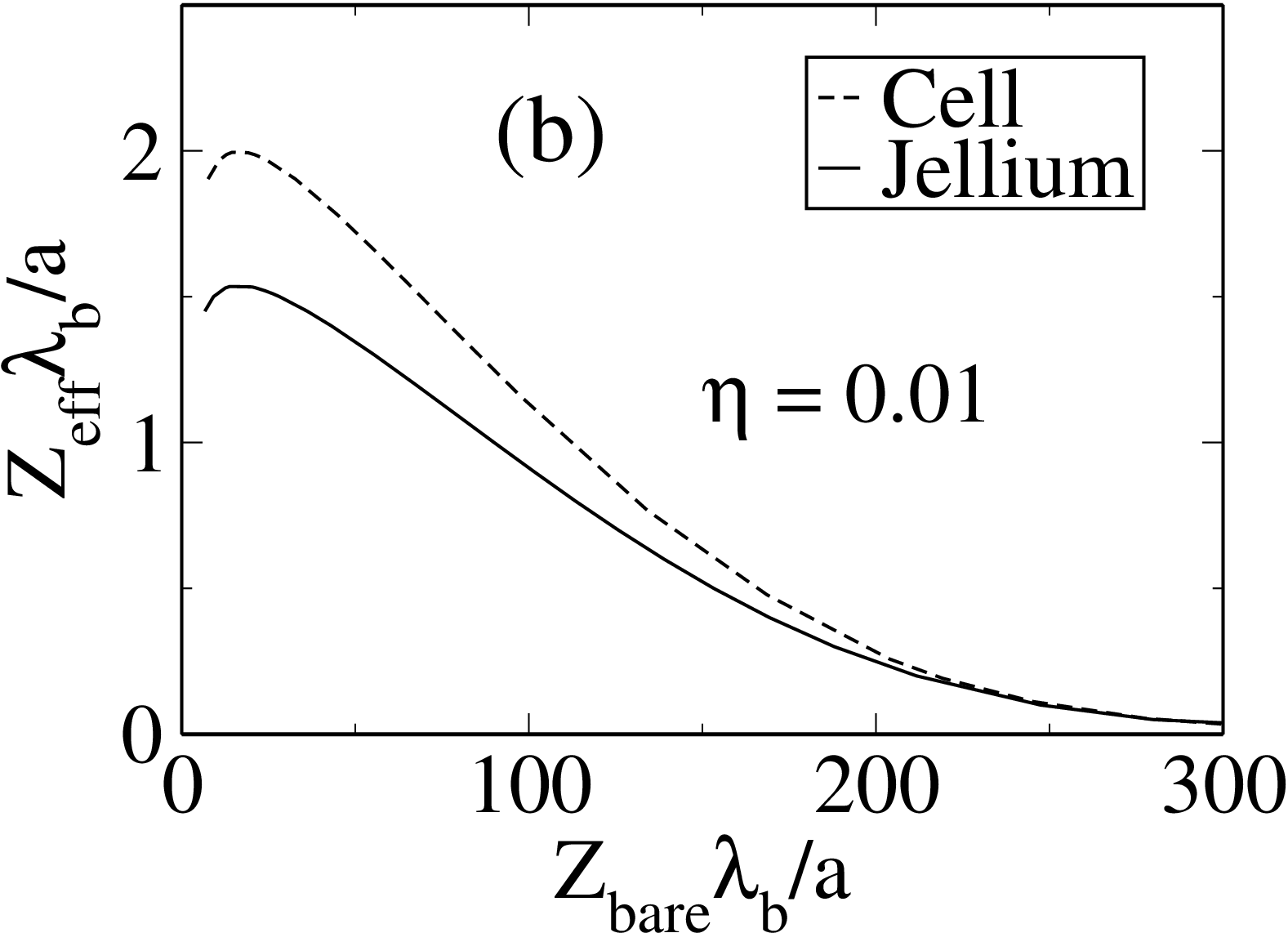}}
\subfigure
{\includegraphics[width=4.5cm,height=4.0cm]{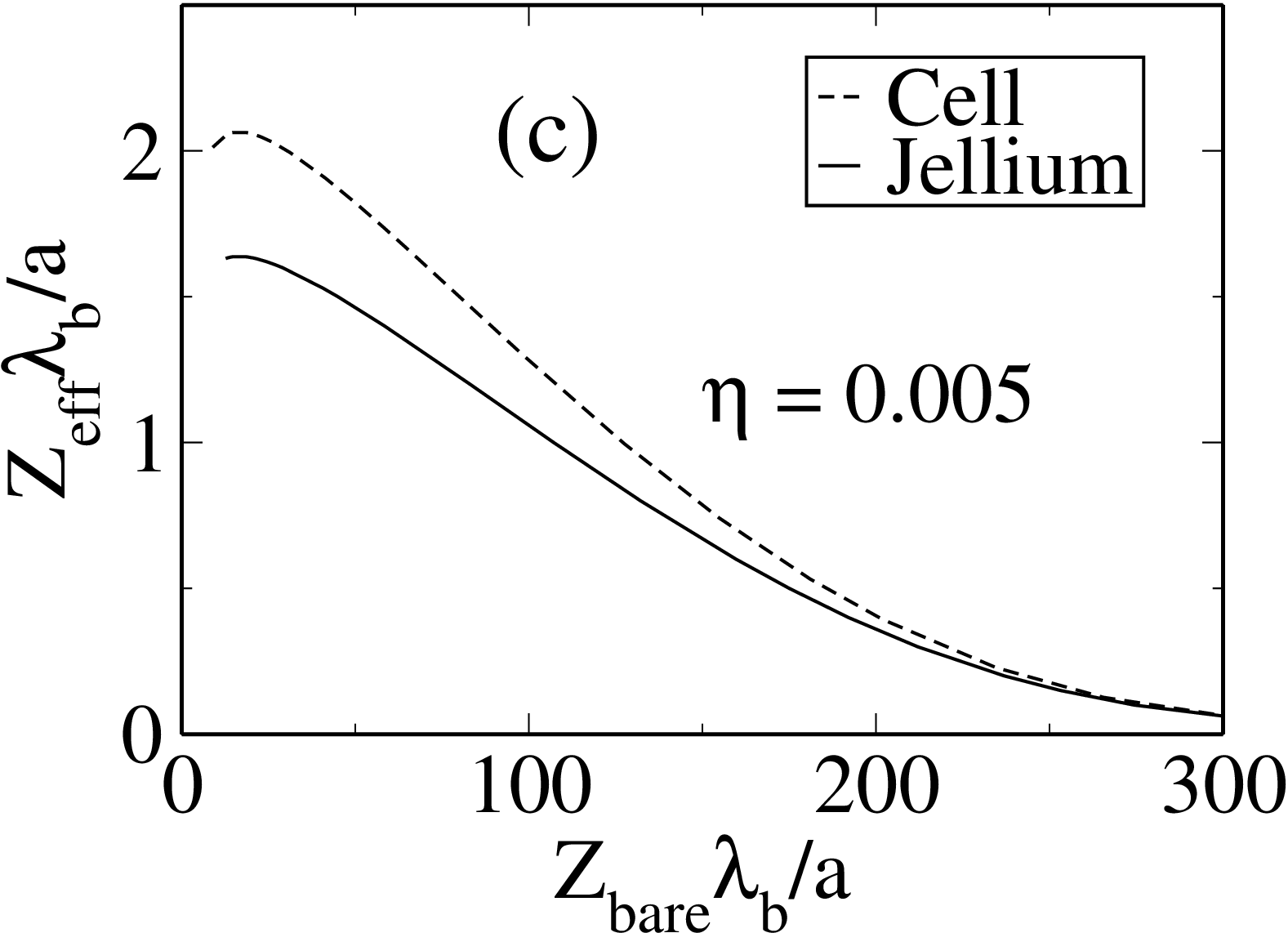} 
\includegraphics[width=4.5cm,height=4.0cm]{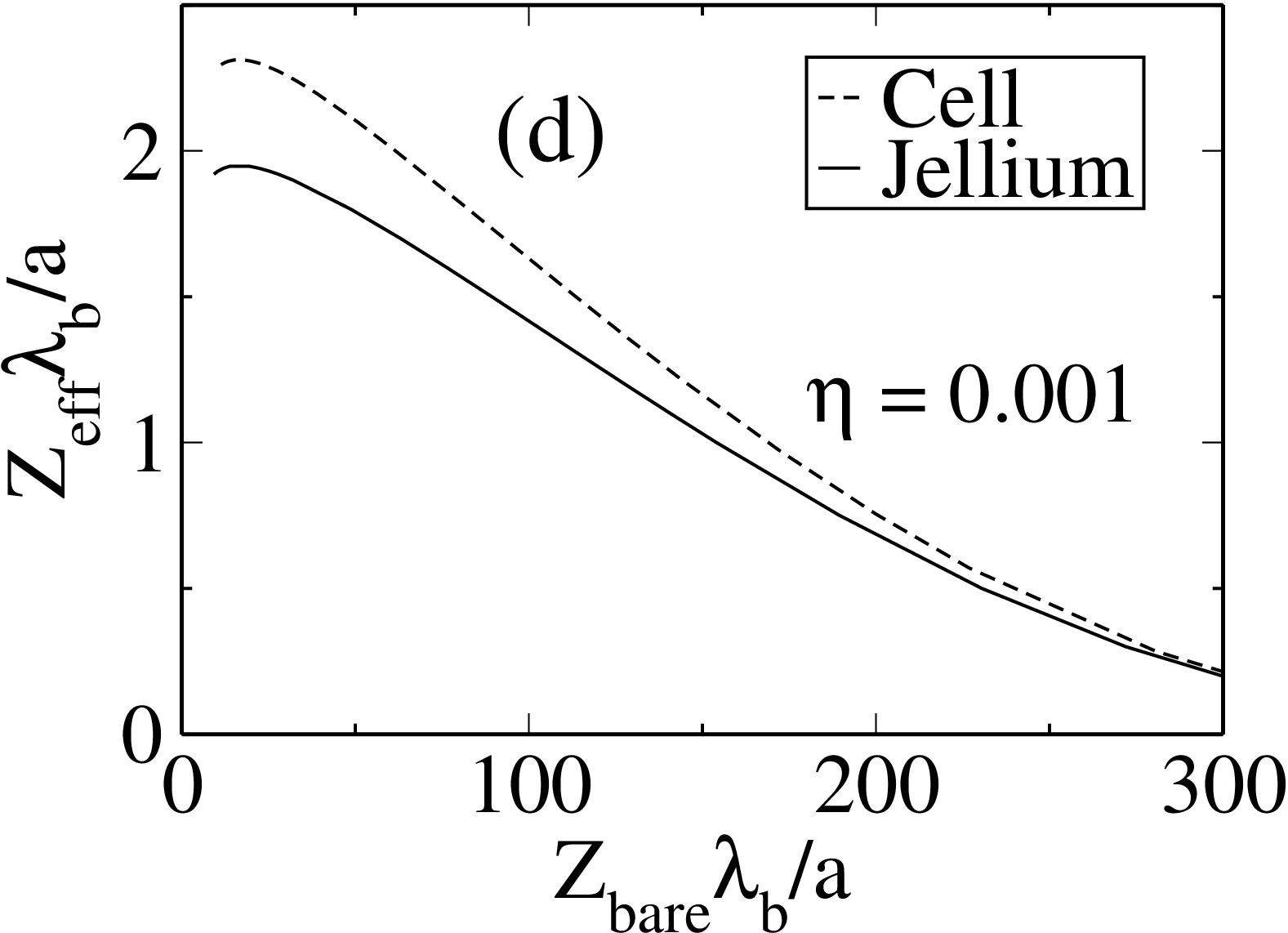}}

\caption{Effective charge as a function of the bare charge for suspensions with trivalent counterions. The volume fractions are a) $\eta=0.125$, b) $\eta=0.01$, c) $\eta=0.005$ and d) $\eta=0.001$. The solid lines have been calculated using the RJM, while the dashed lines have been obtained using the WS cell model \cite{13}. The difference between the two models diminishes as the volume fraction decreases.}
\end{figure}

Fig. 1 compares the effective charge as a function of the bare charge for several colloidal concentrations, using the modified boundary condition within the RJM (solid curves) and within the WS cell model (dashed curves) \cite{13}. In all the calculations, the counterion and the colloidal radii used are $r_{c}=2$ \AA$\,$ and $a=100$ \AA, respectively.

Although both models predict similar qualitative behaviors, there is a significant quantitative difference between the two effective charges.   The effective charges calculated using the RJM lie below the ones calculated in the WS cell approach. A similar result was found for the monovalent ions: apparently,  the colloid-colloid correlations, implicit in the cell model,  reduce the counterion condensation \cite{28}. The difference, however, becomes smaller as the colloidal concentration decreases, as is also observed in the case of 1:1 electrolyte.

Unlike the monovalent situation, the effective charge in the case of trivalent counterions is not a monotonically  increasing function of the bare charge, as can be clearly seen in Fig.1. Instead, after reaching  a maximum,  the effective charge decreases with further increase of the bare charge. This general trend is in perfect agreement with the findings of the Monte Carlo simulations \cite{13}.

\begin{figure}
\centering
\subfigure
{\vspace{0.2cm}
\includegraphics[width=6.0cm,height=4.5cm]{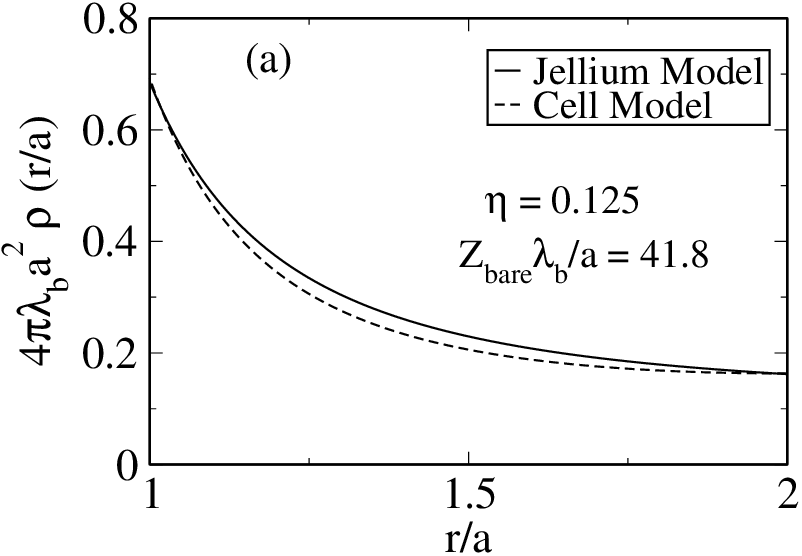}}
\vspace{0.2cm}
\subfigure{\includegraphics[width=6.0cm,height=4.5cm]{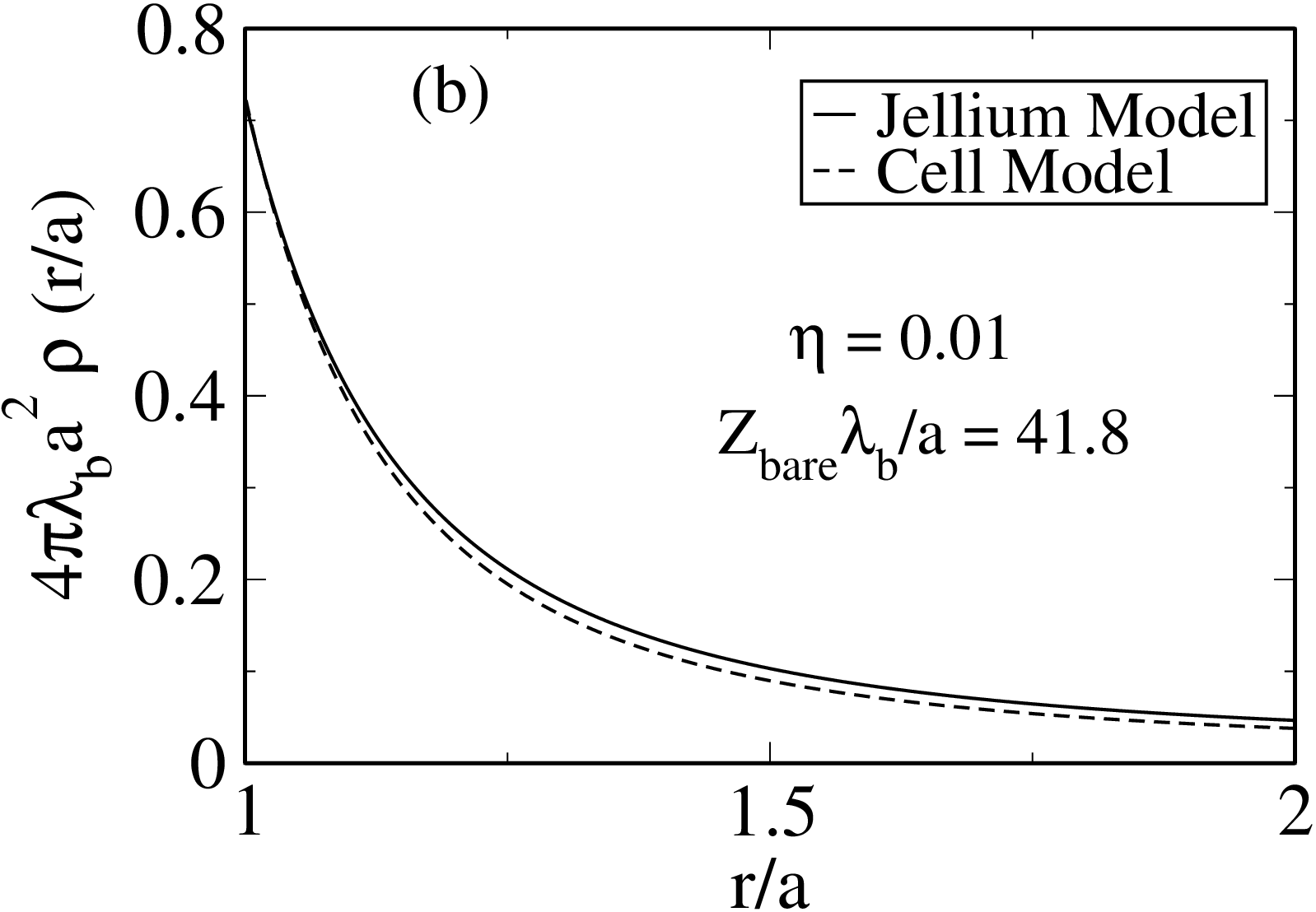}}
\subfigure{\includegraphics[width=6.0cm,height=4.5cm]{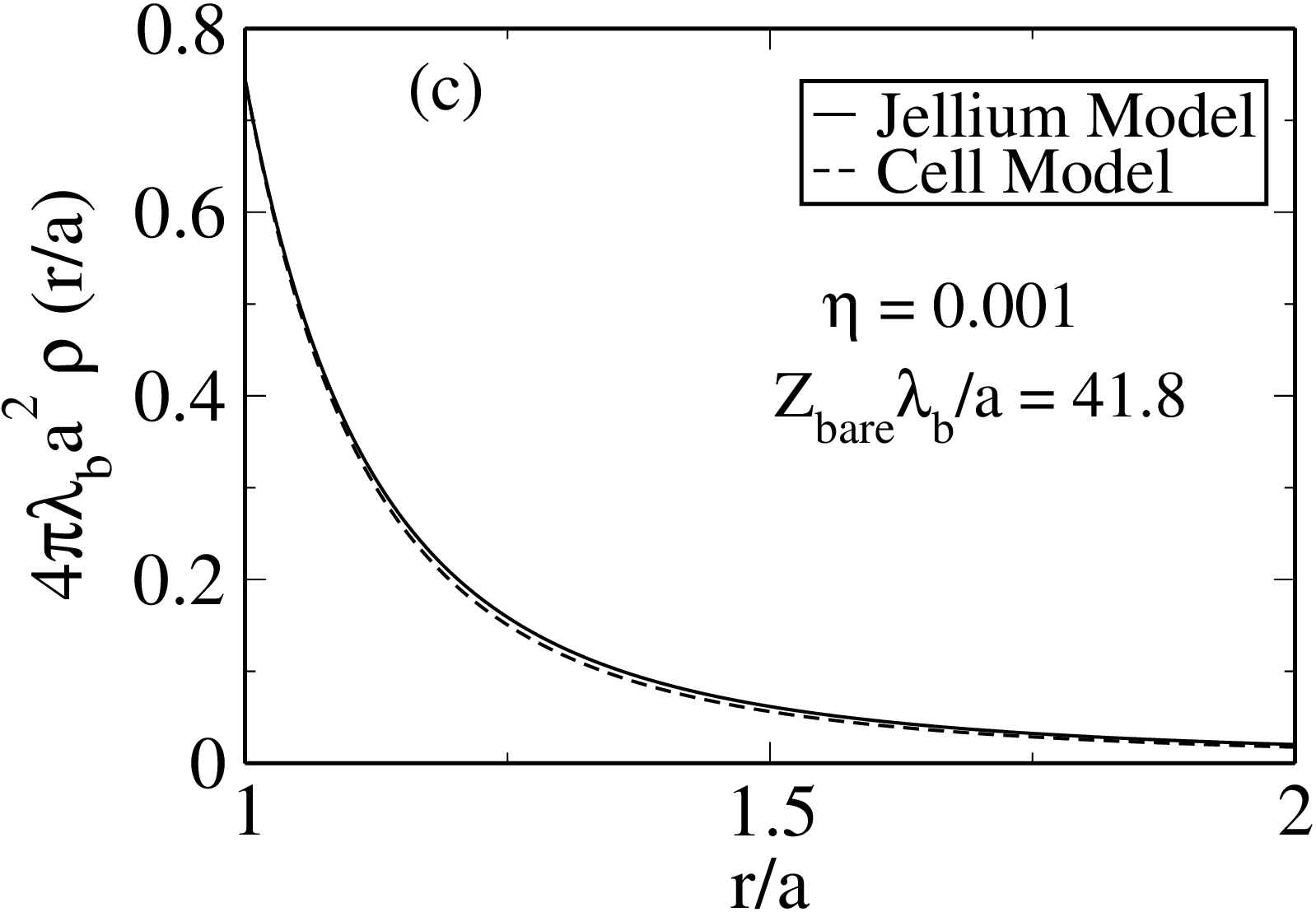}}
\caption{Density profiles calculated using the RJM (solid lines) and the cell model (dashed lines). The volume fractions are a) $\eta=0.125$, b) $\eta=0.01$ and c) $\eta=0.001$. The plasma parameter is $\Gamma=4.413$. We see very small discrepancy, which is further reduced as the volume fraction is decreased.}
\end{figure}

In Fig. 2, the counterion density profiles calculated using both Jellium and WS cell models for a fixed bare charge $Z_{bare}\lambda_{b}/a=41.8$, are displayed for different colloidal concentrations. We conclude that the difference in the density profiles is even less pronounced when compared with the corresponding discrepancy in the effective charges calculated using the two models. It is important to remember that the density profiles obtained using the above theory are only valid after some distance from the colloidal surface. At short distances, the strong-coupling regime dominates over the mean-field.  The present theory
coarse-grains the whole near-field region into a modified boundary condition for the JPB equation.

Besides the effective charges and the ionic distributions, the thermodynamic properties can also be easily calculated in  the framework of the renormalized Jellium and the cell models. The osmotic pressure $P$ within the RJM is a function of the bulk counterion concentration \cite{17,29}:
\begin{equation}
4 \pi \lambda_{b} a^{2} \beta P = \dfrac{3 \eta \lambda_{b}}{a}+\dfrac{{(\kappa a)}^{2}}{\alpha},
\end{equation}
Employing Eq. (9), we have calculated the osmotic pressure as a function of the volume fraction for two fixed bare colloidal charges, $Z_{bare}=41.8$ and $Z_{bare}=100.0$, using both the WS cell model and the RJM. The results are shown in Fig. 3. 
Again, we see only a very small difference at moderate volume fractions. For small volume fractions,  the osmotic pressures are identical within the two models. 

\begin{figure}
 \centering
{\includegraphics[width=8.0cm,height=6cm]{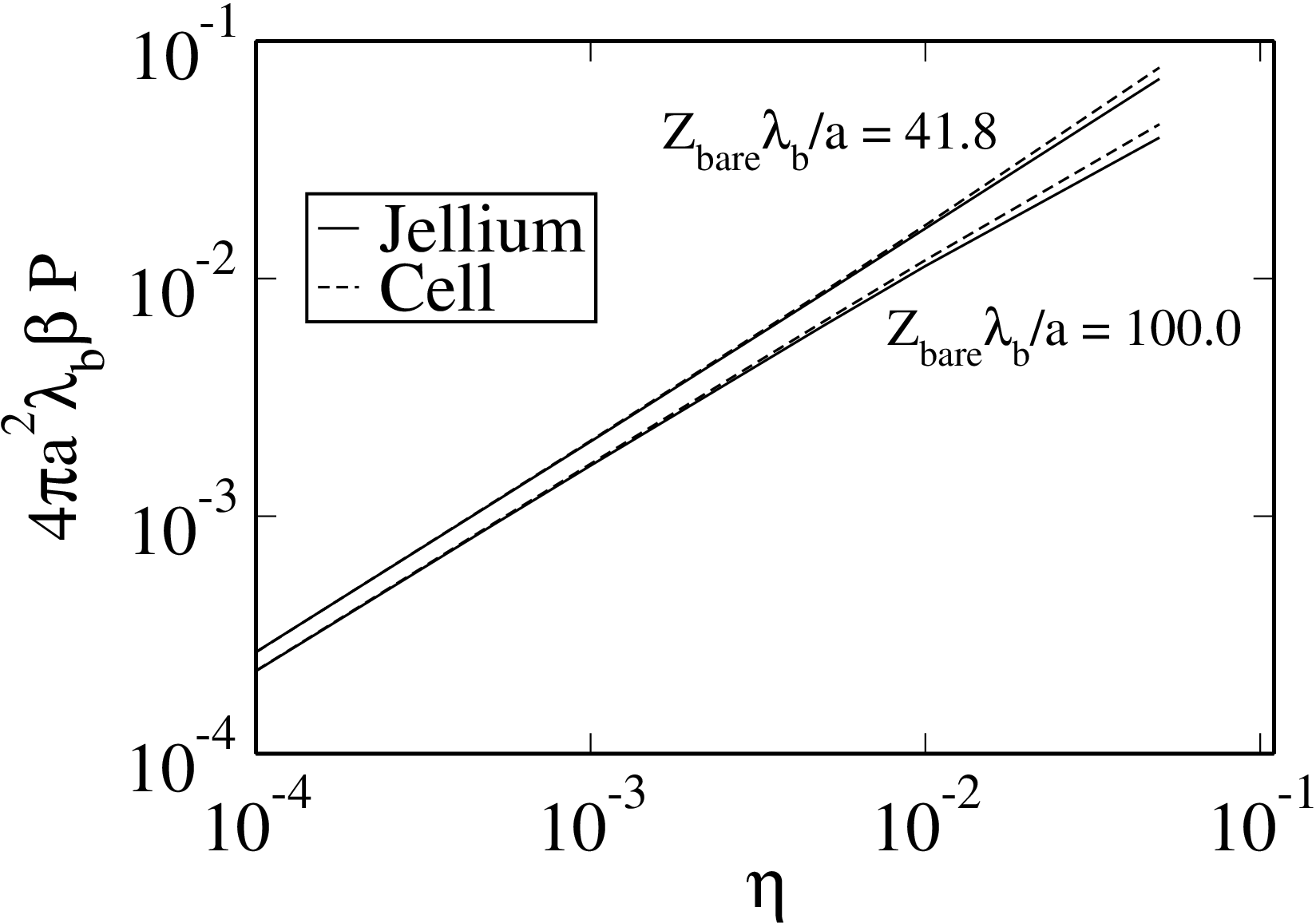}}
\caption{Osmotic pressure as a function of the volume fraction for colloidal suspensions with trivalent counterions and bare charges $Z_{bare}=41.8$ (upper curves) and $Z_{bare}=100.0$ (lower curves). The corresponding plasma parameters are $\Gamma=4.413$ and $\Gamma=6.826$, respectively. The solid lines have been calculated using the RJM, and the dashed lines have been obtained using the WS cell model.}
\end{figure}

\section{Summary and Conclusions}

We have extended the range of applicability of the renormalized Jellium model to describe suspensions with multivalent counterions in the absence of added salt. The model uses Shklovskii's idea to include the counterion correlations as a modified boundary condition for the JPB equation.  A similar strategy has
already proven to be successful for the modified WS PB model \cite{13,14}. Comparing the predictions of the renormalized Jellium model and the WS model,
we find a quantitative difference in the values of the effective charge of colloidal particles.  Since the far field potential within the
renormalized Jellium formalism has precisely the DLVO form, we expect that the effective charges calculated using this formalism 
should be more reliable for structural calculations. On the other hand the measurable thermodynamic quantities such as the osmotic pressure come out to be practically identical in the two models. 

Besides providing an alternative to the cell model, the RJM can be further extended to take into account the colloid-colloid correlations. The homogeneous background charge distribution can be replaced by a non-homogeneous one related to the colloid distribution function. This approach has been successfully implemented for suspensions with monovalent electrolyte \cite{20,21}. Unfortunately, for the case of multivalent counterions, inclusion of colloidal correlations is not so straightforward. To have an accurate structure function one needs to know the interaction potential not only in the far field, but also in the near field. The DLVO effective interaction potential, however, is not valid at short distances where it becomes strongly modified by the counterion correlations \cite{30,31}. The work in this direction is now in progress.

\section{Acknowledgments}

T.E.C. would like to acknowledge useful conversations with Alexandre P. dos Santos. This work was partially supported by 
the CNPq, INCT-FCx, and by the US-AFOSR under the grant FA9550-09-1-0283.

\end{document}